\documentclass[useAMS,usenatbib]{mn2e}

\usepackage{tensor}
\usepackage{multirow}
\usepackage{rotating}
\usepackage{myaasmacros}
\usepackage{epsfig}
\newcommand{\lie}{\mathcal{L}}

\newcommand{\dd}{\mathrm{d}}
\newcommand{\dD}{\mathrm{D}}
\newcommand{\diag}{\mathop{\mathrm{diag}}}
\newcommand{\udot}{\textrm{ }\dot{}\textrm{ }}

\newcommand{\Y}[3]{{\tensor[_{\hspace{1mm}#1\hspace{-1mm}}]{Y}{}_{#2}^{#3}}}
\newcommand{\tens}[1]{{\bf \sf #1}}

\renewcommand{\vec}[1]{\bmath{#1}}

\begin{document}
 
\title[Bianchi Model CMB Polarization]{Bianchi Model CMB Polarization and its Implications for CMB Anomalies}

\author[Andrew Pontzen, Anthony Challinor]{Andrew Pontzen,$^1$\thanks{Email: apontzen@ast.cam.ac.uk} 
  Anthony Challinor$^{1,2}$\thanks{Email: a.d.challinor@ast.cam.ac.uk} \\
  $^1$ Institute of Astronomy, Madingley Road, Cambridge CB3 0HA, UK \\
  $^2$ DAMTP, Centre for Mathematical Sciences, Wilberforce Road, Cambridge CB3 0WA, UK }
\date{Accepted 2007 July 6. Received 2007 June 12.}
\pubyear{2007}

\maketitle

\begin{abstract}
  We derive the CMB radiative transfer equation in the form of a
  multipole hierarchy in the nearly-Friedmann-Robertson-Walker limit
  of homogeneous, but anisotropic, universes classified via their
  Bianchi type. Compared with previous calculations, this allows a
  more sophisticated treatment of recombination, produces predictions
  for the polarization of the radiation, and allows for reionization.
  Our derivation is independent of any assumptions about the dynamical
  behaviour of the field equations, except that it requires
  anisotropies to be small back to recombination; this is already
  demanded by observations.
        
  We calculate the polarization signal in the Bianchi VII$_h$ case,
  with the parameters recently advocated to mimic the several
  large-angle anomalous features observed in the CMB. We find that the
  peak polarization signal is $\sim 1.2\, \mu\mathrm{K}$ for the
  best-fit model to the temperature anisotropies, and is mostly
  confined to multipoles $l<10$.  Remarkably, the predicted
  large-angle $EE$ and $TE$ power spectra in the Bianchi model are
  consistent with WMAP observations that are usually interpreted as
  evidence of early reionization. However, the power in $B$-mode
  polarization is predicted to be similar to the $E$-mode power and
  parity-violating correlations are also predicted by the model; the
  WMAP non-detection of either of these signals casts further strong
  doubts on the veracity of attempts to explain the large-angle
  anomalies with global anisotropy.  On the other hand, given that
  there exist further dynamical degrees of freedom in the VII$_h$
  universes that are yet to be compared with CMB observations, we
  cannot at this time definitively reject the anisotropy explanation.
\end{abstract}


\section{Introduction}\label{sec:introduction}

There are a number of observed features in the large-angle temperature
anisotropies of the cosmic microwave background (CMB) that are
anomalous under the usual assumption of statistically-isotropic,
Gaussian fluctuations (see~\citealt{2007PhRvD..75b3507C} for a recent
summary).  Recent analyses have suggested that a small global
anisotropy in the form described by Bianchi models could be to blame
\citep{2005ApJ...629L...1J}.  In classical general relativity, the
dimension of initial state space leading to models with such
homogeneous anisotropies is always greater than the limitation of
these models to the exactly isotropic case; furthermore, nearly
isotropic models do not necessarily tend to late-time isotropy
\citep{1973ApJ...180..317C}. Although later work has suggested an
inflationary epoch can be responsible for removing anisotropies
\citep{1983PhRvD..28.2118W}, this analysis is not complete
\citep{1999PhRvD..60b3502G}. It is therefore worth investigating
further the observational predictions of such models; whilst to the
skeptic they seem unlikely, the importance of a positive detection
would be great.

Using the CMB models developed in \cite{1973MNRAS.162..307C} and
\cite*{1985MNRAS.213..917B}, based on a specific solution for a
type-VII$_h$ universe with only pressureless matter,
\citet{2005ApJ...629L...1J} showed that known anomalies such as the
low quadrupole amplitude, alignment of low-$l$ modes, large-scale
power asymmetry and the `cold spot' in the CMB could be mimicked. The
work was extended in \cite{2006ApJ...644..701J} to include the
dynamical effect of dark energy, yielding a degeneracy in the
$\Omega_{\Lambda}$ - $\Omega_{\mathrm{M}}$ plane (which effectively
arises through the angular diameter distance relation, since much of
the temperature anisotropy is generated at high redshift). The
degenerate range of parameters able to explain the large-angle
anomalies was shown to be inconsistent with the cosmological
parameters required to explain constraints such as supernovae
observations and the CMB spectrum on smaller scales (i.e.\ the
structure of the acoustic peaks). A more complete statistical analysis
was performed by \cite{2007MNRAS.377.1473B}, in which the
inconsistency of inferred parameters was confirmed.

There are remaining dynamical freedoms in the VII$_h$ model which were
not explored in the above papers. As noted in
\cite{2006ApJ...644..701J}, these will need investigation before any
firm conclusions about the compatibility of the model with
observations can be made. 

A second important issue, hitherto not considered, is whether the
predicted CMB {\it polarization} pattern in the Bianchi model that
best fits the large-angle temperature anisotropies is consistent with
current observations. The reason for its neglect seems to be that
there are currently no full predictions for polarization in Bianchi
models~\citep{2006ApJ...644..701J}. In this paper, we address this
issue by providing a complete and computationally convenient framework
for calculating polarization in these models, and make a first attempt
at confronting the predictions with the three-year data from the
\emph{Wilkinson Microwave Anisotropy Probe}
(\emph{WMAP};~\citealt{2007ApJS..170..335P}).  Polarization of the CMB
in anisotropic cosmologies was first discussed by
\cite{1968ApJ...153L...1R}, where a calculation for Bianchi I models
was outlined. Further consideration has been given to the problem in
\cite{PhysRevD.26.2951}, where Bianchi types V and IX were considered.
There are also a number of papers considering the effect of
homogeneous magnetic fields on polarization
\citep[e.g.][]{1985A&A...151....7M,1987A&A...179...11F}, although we
emphasize that the resulting Faraday rotation of the polarization is
quite distinct from the purely gravitational effect under
consideration here.

This paper is organized as follows.  We briefly review Bianchi models
in Section \ref{sec:preliminaries}, including a more intuitive
derivation of their FRW limit than has previously been
published. In Section \ref{sec:evol-polar-angle} we derive the
zero order evolution of the photon polarization direction along
geodesics, and incorporate these results into a multipole treatment of
the Boltzmann equation that describes the radiative transfer in
Section \ref{sec:boltzmann-heirarchy}.  Using this, we specialize our
model to calculate the temperature and polarization anisotropies
expected in a VII$_h$ model with favoured parameters
\citep{2006ApJ...644..701J} in Section \ref{sec:impl-vii_h}, and
discuss the difficulties of reconciling the results with existing
observations.  We defer a full statistical reanalysis of Bianchi
signatures, including reionization, to future work, where we will also
explore more fully the extra dynamical degrees of freedom in the
models.

\section{Bianchi models}\label{sec:preliminaries}

In this section we give a brief review of the framework that we use
for our calculation. Anisotropic, homogeneous cosmological models can
be classified according to the commutation relations of their spatial
symmetry groups. The most popular classification method is based on
Bianchi's (1897) classification of three-parameter Lie
groups.\nocite{bianchi1897} For the development of these
classifications in the cosmological context see
\cite*{taub1951est,heckmann1962gic,estabrook1968das,EllMac69}, and for
reviews see \cite{wainwright1997dsc} and~\citet{ellis1998cmc}.

We adopt a $-+++$ metric signature, and will use Greek spacetime
indices ($0 \to 3$) for tensor components in a general basis, early
Latin indices ($a$, $b$ etc.\ running from $0 \to 3$) to label the
vectors of any group-invariant tetrad, and middle Latin indices ($i$,
$j$ etc.\ running from $1 \to 3$) to label spatial tetrad vectors in
expressions only involving the spatial vectors. For the latter, we
also use upper-case early Latin ($A$, $B$ etc.) when the tetrad is
time-invariant (see below).  As usual, round brackets denote
symmetrisation on the enclosed indices and square brackets denote
anti-symmetrisation.

A spacetime is said to be homogeneous if it can be foliated into space
sections each admitting at least three linearly-independent
Killing vector fields (KVFs) $\{ \vec{\xi} \}$ (so that
$\lie_{\vec{\xi}}\tens{g}=0$, where
$\lie$ denotes the Lie derivative and $\tens{g}$ is the metric
tensor) with at least one subgroup that acts simply transitively
in the space sections. We denote the elements of the three-dimensional
subgroup by $\vec{\xi}_i$ where $i=1$, $2$, $3$. The commutator of
two KVFs is also Killing and, since the $\vec{\xi}_i$ form
a subgroup, we must have
\begin{equation}
[\vec{\xi}_i,\vec{\xi}_j]  =  C^k_{ij} \vec{\xi}_k .
\end{equation}
The $C^k_{ij} = C^k_{[ij]}$ are the structure constants of the Lie
algebra of the group and are constant in space. (Through the foliation
construction below they may also be shown to be constant throughout
time.) The Jacobi identity,
\begin{equation}
[\vec{\xi}_i , [\vec{\xi}_j,\vec{\xi}_k]]  + 
[\vec{\xi}_j , [\vec{\xi}_k,\vec{\xi}_i]]  + 
[\vec{\xi}_k , [\vec{\xi}_i,\vec{\xi}_j]]  = 0 ,
\end{equation}
restricts the structure constants to satisfy
\begin{equation}
C^m_{n[i} C^n_{jk]} = 0 .
\label{eq:structure}
\end{equation}
Since a constant linear combination of KVFs will also be a KVF, one is
permitted to perform global linear transformations:
\begin{equation}
\vec{\xi}_i \to \vec{\xi}_i'=T_i^j\vec{\xi}_j , \label{eq:lin-tran}
\label{eq:trans}
\end{equation}
under which the structure constants transform as a (mixed) 3-tensor.
Classifying all homogeneous spacetimes amounts to finding solutions
to equation~(\ref{eq:structure}) that are inequivalent under the
linear transformations~(\ref{eq:trans}).

The fiducial classification is achieved by decomposing the structure
constants into irreducible vector and pseudo-tensor parts:
\begin{eqnarray}
C^m_{ij} & = & \epsilon_{ijk}n^{km}+a_i\delta^m_j-a_j\delta^m_i\label{eq:ellmac-decomp} , \nonumber \\
n^{ij}a_i & = &0 \hspace{1cm} \textrm{(Jacobi identities)} ,
\end{eqnarray}
where $n^{ij}$ is symmetric and $\epsilon_{ijk}$ is the alternating
tensor.  Linear transformations can be used to diagonalise $n^{ij}$
and, since $a_i$ is an eigenvector of $n^{ij}$ one may take, without
loss of generality:
\begin{eqnarray}
n^{ij} & = & \diag(n_1,n_2,n_3) , \nonumber \\
a_i & = &(a,0,0) . \label{eq:diagn}
\end{eqnarray}
By rescaling (possibly reversing their directions) and relabelling the
KVFs, we may reduce the structure constants to one of 10
distinct canonical forms that describe the 10 possible group
types (see, for example,~\citealt{EllMac69}).
In these forms, all non-zero $n_i$ and $a$ are either $\pm 1$ except
for the two types with $a n_2 n_3 \neq 0$ in which case there
is an additional parameter $h \equiv a^2/(n_2 n_3)$. 

In order to express homogeneous tensor fields in an invariant way
within this space, one must construct a suitable basis tetrad
$\{\vec{e}_a\}$ that is invariant under the group action, i.e.\
satisfying
\begin{equation}
[\vec{e}_a,\vec{\xi}_i] = 0 . \label{eq:tetrad-build}
\end{equation}
Any homogeneous tensor ($\tens{T}$ where $\lie_{\vec{\xi}}
\tens{T}=0$) has components relative to this tetrad that are constant
throughout space. The unit normal $\vec{n}$ to the hypersurfaces of
homogeneity (satisfying $\vec{n} \cdot \vec{\xi}_i=0$) is necessarily
group-invariant ($[\vec{n},\vec{\xi}_i]=0$) via the Leibniz property
of $\mathcal{L}$, so we can always take $\vec{e}_0 = \vec{n}$. Because
of the foliated construction of the spacetime, one may label the
hypersurfaces with a coordinate time $t$ such that
$\vec{n}_{\alpha}=-t_{,\alpha}$ and hence $\vec{e}_0$ can be shown to
be geodesic.  In any homogeneous hypersurface, we can always construct
the spatial part of the group-invariant tetrad by making an arbitrary
choice for $\vec{e}_i$ at a point, and then dragging this frame out
across the hypersurface with the $\vec{\xi}_i$ (see, for
example,~\citealt{EllMac69}).

In general the elements of the tetrad will not commute so one has
\begin{equation}
[\vec{e}_a,\vec{e}_b] = \gamma_{ab}^c \vec{e}_c ,
\end{equation}
where the $\gamma_{ab}^c = \gamma_{[ab]}^c$ are constant in the
hypersurfaces of homogeneity by the Jacobi identities.
The rotation coefficients encode the covariant derivatives of the tetrad:
\begin{equation}
\Gamma_{cab} \equiv e_c^\alpha e_a^\beta \nabla_\beta e_{b\alpha} ,
\end{equation}
where $e_a^\alpha$ are the components of the tetrad vectors in an
arbitrary basis. The rotation coefficients are related to the
$\gamma_{ab}^c$ by
\begin{eqnarray}
\Gamma_{abc} = \frac{1}{2}\left(-\partial_a g_{bc} + \partial_b g_{ca}
+ \partial_c g_{ab} \right. \nonumber \\
\hspace{3cm} \left. + \gamma_{abc} + \gamma_{cab} - \gamma_{bca}\right) ,
\label{eq:rotation}
\end{eqnarray}
where $g_{ab} \equiv \tens{g}(\vec{e}_a,\vec{e}_b)$ are the tetrad
components of the metric, $\gamma_{abc} \equiv g_{ad} \gamma^d_{bc}$
and $\partial_a$ is the ordinary derivative along the direction of
$\vec{e}_a$. By taking $\vec{e}_0 = \vec{n}$, we have $g_{00} = -1$,
$g_{0i}=0$ and, since $\vec{n}$ is normalized, geodesic and
irrotational, $\gamma^0_{ab} = 0$. The group-invariance of the tetrad
restricts the form of the spatial components $\gamma_{ij}^k$: they are
related to the group structure constants by a linear transformation
and can be classified in the same way (and, necessarily, fall into the
same Bianchi type: see \citeauthor{1973CLP.....6...61M} 1973).

It is always possible to construct the tetrad so that it is
orthonormal; see \cite{EllMac69} for details. However, a convenient
alternative is the {\it time-invariant frame} used
by~\cite{1973MNRAS.162..307C}. In this case, the tetrad is constructed
so that $[\vec{n},\vec{e}_a]=0$. The Jacobi identities applied to
$\vec{n}$, $\vec{e}_a$ and $\vec{\xi}_i$ show that this is still
consistent with the tetrad being group invariant. A specific
construction is to start with a group-invariant tetrad over a
hypersurface and to drag the tetrad along the normal $\vec{n}$ to
cover spacetime. An important consequence is that the $\gamma_{ab}^c$
are constant throughout spacetime for the group-invariant tetrad.
Moreover, $\gamma_{0j}^i=0$ and the only non-vanishing components are
spatial, i.e.\ $\gamma_{ij}^k$. Because the Bianchi classification of
the Killing group and tetrad commutators is the same, it is always
possible to perform a global (three-dimensional) linear transformation
of the time-invariant tetrad to bring the $\gamma_{ij}^k$ equal to the
canonical structure constants, i.e.\ $\gamma_{ij}^k = C_{ij}^k$. We
shall use upper-case early Latin indices ($A$, $B$ etc.) to denote the
spatial elements of this specific time-invariant tetrad.

The spacetime metric can then be written in terms of the
time-invariant tetrad as
\begin{equation}
\tens{g} = -\vec{n} \otimes \vec{n} + g_{AB}(t) \vec{e}^A \otimes
\vec{e}^B ,
\end{equation}
where the basis one-forms $e^A_\alpha = g^{AB} g_{\alpha \beta}
e_B^\beta$ satisfy $e^A_\alpha e_B^\alpha = \delta^A_B$. Here, as usual,
$g^{AB}$ denotes the inverse of $g_{AB}$. The $\vec{e}^A$ are also
group- and time-invariant.
We decompose the
spatial part of the metric following \cite{1968ApJ...151..431M}:
\begin{equation}
g_{AB} = e^{2\alpha(t)}\left(e^{2\beta(t)}\right)_{AB} ,
\end{equation}
where $\beta$ is a matrix with zero trace and the matrix exponential
is defined as usual via Taylor expansion, yielding $\det e^{\beta} =
1$. Thus $\alpha$ represents shape-preserving expansion whilst $\beta$
represents volume-preserving shape deformation.  The expansion rate of
the $\vec{n}$ congruence is $\nabla_\alpha n^\alpha = 3 \dot{\alpha}$
and its shear -- i.e.\ the trace-free, symmetric spatial projection of
$\nabla_\alpha n_\beta$ -- has components
\begin{equation}
\sigma_{AB} = \frac{1}{2} e^{2\alpha} \left(e^{2\beta}\right){}^{\cdot}_{AB} .
\end{equation}
Here, and throughout, overdots denote derivatives with respect to
$t$. We shall make use of conformal time, $\eta$, defined by
$\dd t = e^\alpha \dd \eta$, and we denote derivatives with respect
to $\eta$ with primes.

The field equations are generally more naturally expressed in a
(group-invariant) orthonormal frame, since the metric derivatives then
vanish and one obtains
first order equations for their commutation functions.
The spatial vectors of such a frame will be represented with lower-case
middle Latin indices, $i$, $j$ etc. One can always define a specific
orthonormal frame by~\citep{1969MNRAS.142..129H} 
\begin{eqnarray}
\vec{e}_i = e^{-\alpha} \left(e^{-\beta}\right)_{iA} \vec{e}_A .
\end{eqnarray}
The orthonormal tetrad is completed with $\vec{n}$. The components
of the shear in the orthonormal frame are
\begin{equation}
\sigma_{ij} = (e^{\beta})\udot_{k(i}(e^{-\beta})_{j)k}.
\label{eq:metric-shear}
\end{equation}
Note that this corresponds to the matter shear tensor only in the
case that the fluid flow is not tilted relative to the hypersurfaces;
i.e.\ the case $\vec{n}=\vec{v}$ where $\vec{v}$ is the fluid 4-velocity. In
this paper, we do not specialize in this way.

\subsection{FRW limit}\label{sec:frw-limit}

It is instructive to sketch the derivation of the FRW limit of
spacetimes in the time-invariant frame (the calculation in the
orthonormal frame is described in \citeauthor{EllMac69} 1969 and yields identical
conditions). 

A homogeneous spacetime can only be FRW if the expansion and
3-curvature are isotropic. Vanishing shear requires
$\beta_{AB}=$ const.\ and then we can always set $\beta_{AB}=0$
(i.e.\ diagonalise the spatial metric) with
a constant linear transformation of the $\vec{e}_A$. After the
transformation, the $\gamma_{ij}^k$ will no longer be equal to the
canonical group structure constants (so we denote them with lower-case indices)
but we can perform a further (constant) orthogonal transformation
of the $\vec{e}_i$ (hence preserving $\beta=0$) to bring the
$\gamma_{ij}^k$ to the canonical form up to positive scalings of
$a$, $n_1$, $n_2$ and $n_3$. With $\beta=0$, the 3-curvature is
\begin{eqnarray}
\tensor[^{(3)}]{R}{_{ij}} &=&  \frac{1}{3}\tensor[^{(3)}]{R}{}\delta_{ij}e^{2\alpha} \nonumber \\
&=&  -\frac{1}{4} \left[2\gamma^k_{il} \gamma^l_{jk} +
  2\gamma^l_{ki}\gamma^l_{kj} \right. \nonumber \\
&&\mbox{} \hspace{0.5cm} \left. -\gamma^i_{kl}\gamma^j_{kl} + 2\gamma^k_{kl}(\gamma^{i}_{jl} + \gamma^j_{il})\right]  .
\end{eqnarray}
Zero- and first-order expressions (such
as the one above) will often contain surprising index placement; note
that metric factors are always included where necessary to account for
this.

In terms of the decomposition of $\gamma_{ij}^k$, one obtains the off-diagonal
element $\tensor[^{(3)}]{R}{_{23}}=a(n_2-n_3)$, yielding
the condition
\begin{equation}
a=0\textrm{ or }n_2=n_3
\end{equation}
if the curvature is to be isotropic.
The vanishing of the remainder of the trace-free part further requires
\begin{equation}
n_1^2+n_2n_3 = n_2^2+n_1n_3 = n_3^2 + n_1n_2 ,
\end{equation}
which implies that at least one of the $n_i$ are zero and the remaining
two are equal.
By examining a complete list of distinct group types, these conditions
yield the spaces with an FRW limit (Table \ref{tab:frw-bianchi}). In
each case we can use a volume rescaling (change of $\alpha$) to set
the non-zero $n_i$ to unity or, if the $n_i$ vanish, $a$ to $-1$.
This is enough freedom to set the $\gamma_{ij}^k$ of the time-invariant
frame, with $\beta=0$, equal to the corresponding canonical group
structure constants. For a universe that is close
to FRW, we can therefore treat $\beta$ as a small perturbation
while working in the time-invariant frame with $\gamma_{AB}^C$ still in
canonical form.

\begin{table}
\begin{center}
\begin{tabular}{c|cccc|l}
Type & $a$ & $n_1$ & $n_2$ & $n_3$ & $^{(3)}Re^{2\alpha}$ \\
%for $\beta=0$ \\
\hline
I & 0 & 0 & 0 & 0 & 0 \\
V & $-1$ & 0 & 0 & 0 & $-6$ \\
VII$_0$ & 0 & 0 & 1 & 1 & 0 \\
VII$_h$ & $-\sqrt{h}$ & 0 & 1 & 1 & $-6h$ \\
IX & 0 & 1 & 1 & 1 & $3/2$ \\
\end{tabular}
\caption{Bianchi groups with FRW limit and their structure constants in
  canonical form. In the case VII$_h$, we take $a=-\sqrt{h n_2 n_3}$
  for compatibility with the notation of
  \protect\cite{1973MNRAS.162..307C} and
  \protect\cite{1985MNRAS.213..917B}. Note that $a$ can also be
  positive, but this case will be related by a rotation of the sky
  (constructed by $\vec{e}_1 \to -\vec{e}_1$ followed by $\vec{e}_3
  \to -\vec{e}_3$) and need not be considered separately. However, the
  overall parity must be considered; one way is to consider the
  transformation $\vec{e}_1 \to -\vec{e}_1$ which reverses the sign of
  $a$ and all $n_i$. The final column is the comoving 3-curvature
  scalar in the $\beta=0$ FRW limit.}\label{tab:frw-bianchi}
\end{center}
\end{table}

\section{Photon propagation}\label{sec:evol-polar-angle}

We parameterize the photon propagation direction vector according to the
convention in
\cite{1985MNRAS.213..917B}; although this unusually places the
azimuthal direction along $\vec{e}_1$, it more conveniently reflects
the symmetries of the fiducial Bianchi classification, which
ultimately yields much simpler equations. Thus
in the {\it orthonormal} frame, the photon direction has components
\begin{equation}
p^i =  (\cos\theta,\sin\theta\cos\phi,\sin\theta\sin\phi)
\end{equation}
and the photon 4-momentum is
\begin{equation}
\vec{K} = E (\vec{n} + p^i \vec{e}_i) ,
\end{equation}
where $E$ is the photon energy measured by an observer on a
hypersurface-orthogonal path. It is convenient to introduce the
comoving energy, $\epsilon \equiv E e^\alpha$. In terms of this, the
spatial components of $K$ on the time-invariant tetrad (in
which the geodesic equations take their simplest form) are
\begin{equation}
K_A = \epsilon  \left(e^\beta\right)_{iA} p^i .
\end{equation}

The energy change along a geodesic follows from differentiating
$E = - \vec{K} \cdot \vec{n}$. The geodesic equation then gives the exact
result
\begin{equation}
\epsilon' = - \epsilon e^{\alpha} p^i p^j \sigma_{ij}, \label{eq:epsilon-prime}
\end{equation}
where $\epsilon' \equiv \dd \epsilon / \dd \eta$. This is 
equivalent to the integral result obtained
by~\citet{1969MNRAS.142..129H}. For the evolution of $\theta$ and $\phi$,
we make use of the (exact) geodesic equation in the time-invariant
frame~\citep{1985MNRAS.213..917B}:
\begin{equation}
K_A' = \left(e^{-2\beta}\right)_{BD} \frac{K_C K_D}{\epsilon}
C^C_{BA} .
\label{eq:spatialgoe}
\end{equation}
We only require the evolution of $\theta$ and $\phi$ to zero order in
$\beta$ since the radiation is necessarily isotropic in the FRW limit.
Setting $\beta=0$ in equation~(\ref{eq:spatialgoe}), we find
\begin{eqnarray}
\theta' & = & [a+(n_3-n_2)\cos \phi \sin \phi] \sin \theta  \\
\phi' & = & [n_1-n_3+(n_3-n_2)\cos^2\phi] \cos\theta .
\end{eqnarray}

We denote the (complex) polarization 4-vector by $\vec{P}$. It is normalized
so that $\vec{P}\cdot \vec{P}^* = 1$ and, in the Lorentz gauge, is
orthogonal to $\vec{K}$. The polarization $\vec{P}$ is parallel transported
along the photon geodesic. We are more interested in the observed polarization
$\tilde{\vec{P}}$
(i.e.\ the electric field direction for radiation in a pure state) relative
to $\vec{n}$. This is given by screen-projecting $\vec{P}$ perpendicular to
$\vec{n}$ and the photon direction $\vec{p}$:
\begin{equation}
\tilde{P}^\alpha = \mathcal{H}^\alpha{}_\beta P^\beta ,
\end{equation}
where the screen-projection tensor $\mathcal{H}_{\alpha \beta}$ is defined by
\begin{equation}
\mathcal{H}_{\alpha \beta}
\equiv g_{\alpha \beta} + n_\alpha n_\beta - p_\alpha p_\beta.
\end{equation}
The evolution of $\tilde{\vec{P}}$ follows from the parallel
transport of $\vec{P}$ itself~\citep{2000PhRvD..62d3004C}:
\begin{equation}
\mathcal{H}^\alpha{}_\beta \left( K^\rho \nabla_\rho 
\tilde{P}^\beta \right) = 0 .
\label{eq:ptildetransport}
\end{equation}
We find for the tetrad components of $\tilde{\vec{P}}$ in the
time-invariant basis the exact equation 
\begin{equation}
\dot{\tilde{P}}_A - \tilde{P}^B \sigma_{AB} - p_A \tilde{P}^B p^C \sigma_{BC}
- \dot{\alpha} \tilde{P}_A + p^B \tilde{P}^C \Gamma_{ABC} = 0 ,
\label{eq:tildePdot}
\end{equation}
where $\Gamma_{ABC}$ are the rotation coefficients given by the spatial
components of equation~(\ref{eq:rotation}).\footnote{The metric
derivatives vanish in this case since $g_{IJ}$ is constant in
each hypersurface.}
There is only one degree of freedom in the projected
polarization vector, described by the angle between it and the
$\hat{\vec{e}}_{\theta}$ direction:
\begin{equation}
\vec{\tilde{P}} = \hat{\vec{e}}_{\theta}\cos\psi+\hat{\vec{e}}_{\phi}\sin\psi 
\end{equation}
where $\hat{\vec{e}}_\theta$ and $\hat{\vec{e}}_\phi$
are constructed from the orthonormal-frame vectors. The components
on the orthonormal frame are therefore
\begin{equation}
\tilde{P}^i  =  \left(\begin{array}{c}
-\sin\theta\cos\psi \\
\cos\theta\cos\phi\cos\psi-\sin\phi\sin\psi \\
\cos\theta\sin\phi\cos\psi+ \cos\phi\sin\psi \end{array}\right).
\label{eq:P-dir}
\end{equation}
The polarization is first-order in $\beta$ so we only require the
evolution of $\psi$ at zero order. In this FRW limit,
equation~(\ref{eq:tildePdot}) reduces to
\begin{equation}
2 \psi' =  -n_1-(n_3-n_2)\cos 2\phi .
\end{equation}

Since we are considering models close to FRW, $n_3=n_2$ (see
Table \ref{tab:frw-bianchi}) and our evolution equations
simplify to
\begin{eqnarray}
\theta' & = & -\sqrt{h} \sin \theta  \nonumber \\
\phi' & = & \left(n_1-n_3\right)\cos\theta \nonumber \\ 
2\psi' & = & -n_1 , \label{eq:angle-evols}
\end{eqnarray}
where, for type V universes, one takes $\sqrt{h}=1$.
We see that the polarization only rotates relative to $\hat{\vec{e}}_\theta$
and $\hat{\vec{e}}_\phi$ in type-IX universes.
These general equations for $\theta$ and $\phi$ agree with the
specific cases given in Appendix B of~\citet{1985MNRAS.213..917B}; the
general $\psi$ equation agrees with the previous analysis of type-IX
universes given in~\citet{PhysRevD.26.2951}.

\section{Boltzmann equation}\label{sec:boltzmann-heirarchy}

In this section, we derive the first-order Boltzmann equation
describing polarized radiative transfer in any Bianchi
model with an FRW limit.

CMB polarization is generated by Thomson
scattering and is therefore expected to be only linearly polarized.
We can describe the radiation distribution function in terms of
Stokes parameters $f$, $q$ and $u$ where $f$ gives the expected number
density of photons per proper phase-space volume irrespective of
their polarization state. The parameters $q$ and $u$ describe linear
polarization relative to a basis that we take to be
$\hat{\vec{e}}_\theta$ and $\hat{\vec{e}}_\phi$ for propagation along $p^i$.
Then $(f+q)/2$ is the expected
phase-space number density of photons that would be found in
the $\hat{\vec{e}}_\theta$ linear-polarization state, and $(f+u)/2$
is the same for the state rotated by $45\deg$ (towards $\hat{\vec{e}}_\phi$).
In the absence of scattering, $f$ is conserved along the photon
path in phase space and $q$ and $u$ are conserved if referred to bases
that rotate like $\tilde{\vec{P}}$ in equation~(\ref{eq:ptildetransport}).
We parameterize $f$, $q$ and $u$ by the photon-direction angles
$\theta$ and $\phi$ and by comoving energy $\epsilon$. Since
these parameters are defined relative to a group-invariant tetrad,
$f$ is independent of position in the hypersurfaces of homogeneity. The
same is true for the Stokes parameters since the basis on which they are
defined is constructed in an invariant manner.

The Lagrangian derivative of $f$ in phase space is
\begin{equation}
\frac{\dD f}{\dD \eta} = \frac{\partial f}{\partial \eta}
+ \frac{\partial f}{\partial \theta} \theta' +
\frac{\partial f}{\partial \phi} \phi' +
\frac{\partial f}{\partial \epsilon} \epsilon',
\label{eq:deriv}
\end{equation}
and is only non-zero because of scattering. The quantities
$\epsilon'$, $\theta'$ and $\phi'$ are given by equations
(\ref{eq:epsilon-prime}) and (\ref{eq:angle-evols}).  Anisotropies are
formed through the last term in equation~(\ref{eq:deriv}) and by
Thomson scattering off electrons with a non-zero peculiar velocity;
both effects are first order. The remaining terms `advect' the
resulting pattern on the sphere using the zero-order transport
equations. The energy dependence of the anisotropies is therefore
proportional to $\epsilon \partial \bar{f} / \partial \epsilon$ where
$\bar{f}$ is the Planck distribution function in the FRW background.
We define the dimensionless temperature anisotropies
$\Theta(\theta,\phi;\eta)$ by
\begin{equation}
f(\vec{K};\eta) = \bar{f}(\epsilon)\left(1- \frac{\dd \ln \bar{f}}{\dd \ln \epsilon}
\Theta \right) .
\end{equation}
Polarization is generated by scattering
the anisotropies and, since Thomson scattering is achromatic, the polarization
has the same $\epsilon$ spectrum as the anisotropies. We can therefore
introduce dimensionless `thermodynamic-equivalent'
Stokes parameters $Q(\theta,\phi;\eta)$ and $U(\theta,\phi;\eta)$
as
\begin{equation}
(q\pm i u)(\vec{K};\eta) = - \frac{\dd \bar{f}}{\dd \ln \epsilon}
(Q\pm i U) .
\end{equation}
The quantities $Q\pm i U$ have spin-weight $2$, i.e.\ under a change
of basis 
\begin{equation}
\hat{\vec{e}}_\theta + i \hat{\vec{e}}_\phi \to e^{i\chi} (\hat{\vec{e}}_\theta +
i \hat{\vec{e}}_\phi) \quad \Rightarrow \quad
Q \pm i U \to  e^{\pm 2 i\chi}(Q \pm i U) .
\end{equation}

Expressed in this way, the time dependence of the temperature anisotropies
obeys:
\begin{equation}
\frac{\partial \Theta}{\partial \eta} = \frac{\dD \Theta}{\dD \eta} - \frac{\partial \Theta}{\partial \theta}\theta' - \frac{\partial \Theta}{\partial \theta}\phi' - e^{\alpha} p^i p^j \sigma_{ij} , 
\end{equation}
where the first term on the right describes the Thomson scattering
kernel (Section~\ref{sec:scattering-equations}), the next two describe
advection of the patterns on the sphere, and the final term is
gravitational redshifting in the anisotropic expansion due to the shear.
The advection terms are more transparent when interpreted as being due to
the spatial dependence (relative to a parallel-propagated rather than
time-invariant basis) transforming to angular dependence through free
streaming. For polarization, we have
\begin{eqnarray}
\frac{\partial (Q \pm iU)}{\partial \eta} & = &\frac{\dD (Q\pm iU)}{\dD \eta}  -  \frac{\partial (Q \pm iU)}{\partial \theta}\theta' - \frac{\partial (Q \pm iU)}{\partial \phi}\phi' \nonumber \\ 
& &  \pm 2i ( Q \pm iU) \psi' ,
\end{eqnarray}
where the last term arises from polarization rotation. Again, the first term
on the right describes Thomson scattering (see
Section~\ref{sec:scattering-equations}).

We expand the temperature anisotropies in terms of spherical harmonics
about the propagation direction:
\begin{equation}
\Theta(\vec{p}) = \sum_{lm} \Theta_l^m Y_l^m(\vec{p}) \textrm{ .}
\end{equation}
For polarization, we expand in spin-weighted spherical
harmonics\footnote{For a brief review, see Appendices A-C
of~\citet*{2002PhRvD..65b3505L}.} as
\begin{equation}
(Q \pm iU)(\vec{p}) = \sum_{lm} (E_l^m \pm i B_l^m) \Y{\pm 2}{l}{m}(\vec{p}) .
\end{equation}
Note that the Bianchi models are not parity invariant; the mirror
universe can be obtained using the standard transformations
\begin{eqnarray}
\Theta_l^m & \to & (-1)^l\Theta_l^m \nonumber \\
E_l^m & \to & (-1)^l E_l^m \nonumber \\
B_l^m  & \to &  (-1)^{(l+1)} B_l^m\label{eq:parity} \textrm{ ,}
\end{eqnarray}
or, equivalently, flipping the signs of $a$ and all $n_i$ (which
inverts the direction of the $\vec{e}_1$ axis).

\subsection{Gravitational redshifting}

The effect of the shear on the observed temperature pattern reads
\begin{equation}
\left(\frac{\partial \Theta}{\partial \eta}\right)_{\mathrm{shear}} =
-e^{\alpha} p^i p^j\sigma_{ij} , \label{eq:shear}
\end{equation}
which is a pure quadrupole. In terms of spherical harmonics,
\begin{eqnarray}
\left(\frac{\partial \Theta_2^{2}}{\partial \eta}\right)_{\mathrm{shear}} & = & \sqrt{\frac{2\pi}{15}} \left(\sigma_{33}-\sigma_{22}+2i\sigma_{23}\right)e^{\alpha} \nonumber \\
\left(\frac{\partial \Theta_2^{1}}{\partial \eta}\right)_{\mathrm{shear}} & = & \sqrt{\frac{8\pi}{15}} \left(\sigma_{12}-i\sigma_{13}\right)e^{\alpha} \nonumber \\
\left(\frac{\partial \Theta_2^{0}}{\partial \eta}\right)_{\mathrm{shear}} & = & -\sqrt{\frac{4\pi}{5}} \sigma_{11} e^{\alpha} \nonumber \\
\left(\frac{\partial \Theta_2^{-m}}{\partial \eta}\right)_{\mathrm{shear}} & = & (-1)^m\left(\frac{\partial \Theta_2^{m}}{\partial \eta}\right)^*_{\mathrm{shear}} \textrm{ ,}\label{eq:shear-harmonics}
\end{eqnarray}
where $\sigma_{ij}$ are the components of the shear in the orthonormal
frame. We see that shear injects power at $l=2$, but subsequent advection
generally transports this to higher $l$.

\subsection{Advection equations}

The advection part of the Boltzmann equation for the temperature is
\begin{equation}
\left(\frac{\partial \Theta(\theta,\phi)}{\partial \eta}\right)_{\mathrm{advec}} \equiv 
- \frac{\partial \Theta(\theta,\phi)}{\partial \theta}\theta' - \frac{\partial \Theta(\theta,\phi)}{\partial \phi}\phi' ,
\end{equation}
with $\theta'$ and $\phi'$ given by
equations~(\ref{eq:angle-evols}). Making use of the results
\begin{eqnarray}
\sin\theta \frac{\partial {}_s Y_l^m}{\partial \theta} &=&
\frac{l}{l+1}\sqrt{\frac{[(l+1)^2-m^2][(l+1)^2-s^2]}{
(2l+1)(2l+3)}} {}_s Y_{l+1}^m \nonumber \\
&& \hspace{-1.5cm} + \frac{ms}{l(l+1)} {}_s Y_l^m
- \frac{l+1}{l}\sqrt{\frac{(l^2-m^2)(l^2-s^2)}{(2l-1)(2l+1)}} {}_s Y_{l-1}^m 
\nonumber \\
\cos\theta {}_s Y_l^m &=& \frac{1}{l+1}
\sqrt{\frac{[(l+1)^2-m^2][(l+1)^2-s^2]}{(2l+1)(2l+3)}}
{}_s Y_{l+1}^m \nonumber \\
&& \hspace{-1.5cm} - \frac{ms}{l(l+1)} {}_s Y_l^m
+ \frac{1}{l}\sqrt{\frac{(l^2-m^2)(l^2-s^2)}{(2l-1)(2l+1)}} {}_s Y_{l-1}^m 
\end{eqnarray}
which follow from standard recursion relations for the related
Wigner functions $D^l_{-m s}(\phi,\theta,0)$
(e.g.~\citealt*{1960quthanmo.book.....V}), we find for the advective
contribution to the time derivative of the multipoles 
\begin{equation}
\left(\frac{\partial \Theta_l^m}{\partial \eta}\right)_{\mathrm{advec}} =
\sum_{l'=l-1}^{l+1} f_{ll'}^m \Theta_{l'}^m .\label{eq:temp-adv-sum}
\end{equation}
Here,
\begin{eqnarray}
f_{l,l-1}^{m} & = & \sqrt{\frac{l^2-m^2}{(2l-1)(2l+1)}}\left[i m \Delta n
+ (l-1)\sqrt{h}\right] \nonumber \\
f_{l,l}^{m} & = & 0 \nonumber \\
f_{l,l+1}^{m} & = & \sqrt{\frac{(l+1)^2-m^2}{(2l+1)(2l+3)}} \left[im
\Delta n -(l+2)\sqrt{h}\right]  \label{eq:temp-adv-terms}
\end{eqnarray}
with $\Delta n \equiv n_3 -n_1$. Note that ${}_\pm f_{ll'}^{-m} =
{}_\mp f_{ll'}^{m}{}^*$, as required by the reality of
$\Theta(\theta,\phi)$ in equation~(\ref{eq:temp-adv-sum}).

For polarization, we have
\begin{eqnarray}
\left(\frac{\partial (Q\pm i U)}{\partial \eta}\right)_{\mathrm{advec}}
&\equiv &
- \frac{\partial (Q\pm i U)}{\partial \theta}\theta' - \frac{\partial (Q\pm i U)}{\partial \phi}\phi' \nonumber \\
& & \pm 2i \psi' (Q\pm i U) .
\end{eqnarray}
We write
\begin{equation}
\left(\frac{\partial(E_l^m \pm iB_l^m)}{\partial \eta}\right)_{\mathrm{advec}}=
\sum_{l'=l-1}^{l+1} \tensor[_{\pm}]{g}{}_{ll'}^{m}(E_{l'}^m \pm iB_{l'}^m)
,
\label{eq:ebadvec}
\end{equation}
which may be expressed
\begin{eqnarray}
\left(\frac{\partial E_l^m}{\partial \eta}\right)_{\mathrm{advec}} & = & \frac{1}{2} \sum_{l'=l-1}^{l+1} \left[\left(\tensor[_+]{g}{}_{ll'}^{m}+\tensor[_-]{g}{}_{ll'}^{m}\right)E_{l'}^m \right. \nonumber \\
&&\hspace{0.7cm} \left. +i\left(\tensor[_+]{g}{}_{ll'}^{m}-\tensor[_-]{g}{}_{ll'}^{m}\right)B_{l'}^m \right] , \nonumber \\
\left(\frac{\partial B_l^m}{\partial \eta}\right)_{\mathrm{advec}} & = & \frac{1}{2} \sum_{l'=l-1}^{l+1} \left[\left(\tensor[_+]{g}{}_{ll'}^{m}+\tensor[_-]{g}{}_{ll'}^{m}\right)B_{l'}^m \right. \nonumber \\
&&\hspace{0.7cm} \left. -i\left(\tensor[_+]{g}{}_{ll'}^{m}-\tensor[_-]{g}{}_{ll'}^{m}\right)E_{l'}^m \right]. \label{eq:g-defn}
\end{eqnarray}
We see that there are two modes of propagation;
one transfers power amongst $l$ and the other mixes $E$ and $B$-modes.
The mixing terms are
\begin{eqnarray}
\tensor[_+]{g}{}_{l,l}^{m}-\tensor[_-]{g}{}_{l,l}^{m} & = & -2i n_1
+ \frac{4 \left(\sqrt{h}-i m\Delta n\right) m}{l (l+1)} , \nonumber \\
\tensor[_+]{g}{}_{l,l+1}^{m}-\tensor[_-]{g}{}_{l,l+1}^{m} & = & \tensor[_+]{g}{}_{l,l-1}^{m}-\tensor[_-]{g}{}_{l,l-1}^{m} =0 .\label{eq:mix-terms}
\end{eqnarray}
Polarization is generated from Thomson scattering as a pure electric
quadrupole (see Section~\ref{sec:scattering-equations}) but $B$-modes
can subsequently be produced through advection. This happens in
all nearly-FRW Bianchi models except type I.
The $n_1$ term in equations~(\ref{eq:mix-terms}) arises from the
polarization rotation $\psi'$; the remaining terms are from the evolution
of the photon direction relative to the invariant frame. For the
power transfer, we find
\begin{eqnarray}
\tensor[_+]{g}{}_{l,l}^{m}+\tensor[_-]{g}{}_{l,l}^{m} & = & 0 ,\nonumber \\
\tensor[_+]{g}{}_{l,l-1}^{m}+\tensor[_-]{g}{}_{l,l-1}^{m} & = & 2\sqrt{\frac{(l^2-m^2)(l^2-4)}{l^2(2l-1)(2l+1)}}\times \nonumber \\
& & \hspace{1cm}\left[(l-1)\sqrt{h} + im\Delta n \right] , \nonumber \\
\tensor[_+]{g}{}_{l,l+1}^{m}+\tensor[_-]{g}{}_{l,l+1}^{m} & = & 2\sqrt{\frac{\left[(l+1)^2-m^2\right]\left[(l+1)^2-4\right]}{(l+1)^2(2l+1)(2l+3)}}\times
\nonumber \\
& & \hspace{1cm}\left[-(l+2)\sqrt{h}+im\Delta n\right] ,
\label{eq:pol-tran-terms}
\end{eqnarray}
which are not affected by polarization rotation. Note that
\begin{equation}
{}_\pm g_{ll'}^{-m} = {}_\mp g_{ll'}^{m}{}^* ,
\end{equation}
as required by the reality of $E(\theta,\phi)$ and $B(\theta,\phi)$ in
equation~(\ref{eq:ebadvec}).

\subsection{Scattering equations}\label{sec:scattering-equations}

We use a standard Thomson scattering kernel in the form derived
by~\citet{1997PhRvD..56..596H} \citep[see also ][]{1978AN....299...13D}:
\begin{eqnarray}
\left(\frac{D (E_l^m \pm iB_l^m)}{D \eta}\right)
&=&\tau' \biggl( -(E_l^m \pm iB_l^m) \nonumber \\
& & + \frac{3}{5}\delta_{l2}(E_2^m-\frac{1}{\sqrt{6}}\Theta_2^m)\biggr) ,
\nonumber \\
\left(\frac{D \Theta_l^m}{D \eta}\right) &=&
\tau'\biggl(-\Theta_l^m(1-\delta_{l0}) \nonumber \\
& & \hspace{-1.5cm} +\frac{1}{10}\delta_{l2}(\Theta_2^m-\sqrt{6}E_2^m) + \delta_{l1}
\tilde{u}_m \biggr) , \label{eq:scatter}
\end{eqnarray}
where $\tau'=n_e \sigma_{t} e^{\alpha}$ gives the scattering rate in
conformal time, and
\begin{eqnarray}
\tilde{u}_{-1} & = & \sqrt{\frac{2\pi}{3}}(u_2+iu_3) \nonumber \\
\tilde{u}_{0} & = & \sqrt{\frac{4\pi}{3}} u_1 \nonumber \\
\tilde{u}_{1} & = & \sqrt{\frac{2\pi}{3}}(-u_2+iu_3)\label{eq:silly-u}
\end{eqnarray}
are the dipole moments of the electron peculiar velocity in the orthonormal
frame. It follows from equation~(\ref{eq:scatter}) that Thomson scattering
of the temperature quadrupole generates polarization that is an
$E$-mode quadrupole.

\section{Implementation for {VII$_h$} universes}\label{sec:impl-vii_h}

\subsection{Field equations}\label{sec:field-equations}

The complete set of field equations are available from, for example,
\cite{wainwright1997dsc}. Naturally, these reduce at zeroth-order to
the standard Friedmann and acceleration equations, so that
\begin{eqnarray}
e^{\alpha_0}\frac{\dd \eta}{\dd z} & = & \frac{-1}{H(z)} \nonumber \\
& = & -H_0^{-1}\left(\Omega_{\Lambda,0}+\Omega_{K,0}(1+z)^2 \right. \nonumber \\ 
& & \left. \hspace{2cm} + \Omega_{\mathrm{M},0}(1+z)^3\right)^{-1/2} , \label{eq:friedmann}
\end{eqnarray}
where $\Omega_{\mathrm{M},0}$, $\Omega_{\Lambda,0}$ and $\Omega_{K,0}$ have
their usual meanings, $(1+z)^{-1}=e^{\alpha-\alpha_0}$ and $H=\dot{\alpha}$.
The Friedmann constraint equation relates the
group parameter $h$ to the curvature (see
Table~\ref{tab:frw-bianchi}): $h = H_0^2 e^{2\alpha_0} \Omega_{K,0}$.

The evolution of the shear is required at first order and is
provided by the trace-free part of the spatial evolution equations,
\begin{eqnarray}
\dot{\sigma}_{ij}=-3H\sigma_{ij} - \tensor[^{(3)}]{S}{_{ij}}\label{eq:shear-decay}
\end{eqnarray}
in the orthonormal frame. Here,
\begin{equation}
^{(3)}S_{ij}\equiv \tensor[^{(3)}]{R}{_{ij}}-\tensor[^{(3)}]{R\delta}{_{ij}}/3
\end{equation}
is the trace-free part of the intrinsic 3-curvature of the homogeneous
hypersurfaces. Also, we have assumed a perfect fluid (i.e.\ zero
anisotropic stress) and the shear equation~(\ref{eq:shear-decay})
holds only at first order.  It is convenient that, in the VII$_h$
case, $^{(3)}S_{12}$ and $^{(3)}S_{13}$ are zero to this accuracy,
and, furthermore, no coefficient of $\beta_{12}$ or $\beta_{13}$
enters into the expression for $^{(3)}S_{ij}$, so that one may study a
simple model in which $\sigma_{ij}=0$ except for
$\sigma_{12},\sigma_{13} \propto e^{-3\alpha}$. The linear constraint
equations show that the matter in such a model contains vorticity,
i.e.\ the separation of neighbouring particles rotates relative to
inertial gyroscopes.

Whilst it is not prohibitively difficult to implement a numerical
solution for the most general case, we defer such a treatment to a
later paper. Instead, we take advantage of the simplified solutions to
derive the polarization in the favoured models of
\cite{2006ApJ...644..701J}.

We shall adopt the standard assumption that the CMB signal from
global anisotropy adds linearly to that from inhomogeneities. This is
clearly correct insofar as the linear-order perturbations are
concerned; however, given that generic anisotropy modes grow towards
the initial singularity, there is no guarantee that standard
inflationary mechanisms for generating inhomogeneities can be invoked.
Ignoring this potential inconsistency is pragmatic, but investigation
would certainly be necessary if the resulting models gain any
significant observational support.

\subsection{Tilt decay}\label{sec:tilt-decay}

To be consistent with our assumption of small departures from FRW
symmetry, we assume that all peculiar velocities are small.  If we
write the total momentum density of all matter and radiation as
$\vec{P}^{(\mathrm{tot})} = \sum_n
(\rho_{(n)}+p_{(n)})\vec{u}_{(n)}$,  the linear constraint equation
relates the spatial components to the shear:
\begin{equation}
8 \pi P^{\mathrm{(tot)}}_i = 
e^{-\alpha}(\sigma_{jk} C^j_{ki} - \sigma_{ij} C^k_{kj})
\label{eq:u-constraint}
\end{equation}
to first order in the orthonormal frame.
Here, the $C^i_{jk}$ are the structure constants in canonical form.
In the restricted solution $\sigma_{12}, \sigma_{13} \propto
e^{-3\alpha}$, so, assuming the total momentum density is dominated by
a barotropic fluid with equation of state $p=w\rho$ for constant $w$, we have
$\rho \propto e^{-3(1+w)\alpha}$ and
\begin{eqnarray}
|\vec{u}| = (u_i u^i)^{1/2} & \propto & e^{(3w-1)\alpha} \nonumber \\
& \propto & \left\{ \begin{array}{ll}
\mathrm{constant} & w=1/3 \\
e^{-\alpha} & w=0 . \end{array}\right.\label{eq:tilt-decay}
\end{eqnarray}

This behaviour of $|\vec{u}|$ is consistent with momentum
conservation. To see this, consider the Euler equation for a
non-interacting ideal fluid in the time-invariant frame.  The fluid
pressure is constant on surfaces of homogeneity but gradients
proportional to $\dot{p} u_A$ appear in the fluid rest frame. These
accelerate the fluid so that
\begin{eqnarray}
\dot{p}u_A + (\rho+p)\dot{u}_A & =& 0 .\label{eq:vel-euler}
\end{eqnarray}
Solving gives $u_A \propto e^{3w\alpha}$ and, recalling the zero-order
metric $g^{AB} = e^{-2\alpha}$, we recover equation~(\ref{eq:tilt-decay}).

The above introduces a complication in multi-fluid models, which
appears to have been overlooked in recent work. For two components,
say, the tilt velocities $\vec{u}_{(1)}, \vec{u}_{(2)}$ need not be
the same (except in the case of strong coupling).  Only the total
momentum density,
$(\rho_1+p_1)\vec{u}_{(1)}+(\rho_2+p_2)\vec{u}_{(2)}$ is constrained
by the shear so there is additional freedom in the
solution.\footnote{We note that, with this effect in mind, the
  application of the term `universal vorticity' to describe VII$_h$
  cosmologies is an oversimplification, since the vorticity of the
  dark matter need bear no resemblance to that of the baryons.}
Although the dark matter density will be dominant around the time of
recombination, $z \sim z_{\mathrm{LSS}}$, the tilt velocity of
relevance for the CMB is manifestly that of the baryons.  Given that
the baryons are tightly coupled to the photons until the last
scattering surface, they experience a significant pressure and their
tilt decay will be halted; this will not be the case for the dark
matter.  Thus, one needs to consider with care how to estimate the
electron velocity in the Thomson scattering terms~(\ref{eq:silly-u}).

For most of cosmic history before recombination, the baryon--photon
plasma has an equation of state parameter close to $w = 1/3$.
A simple approximation is thus obtained by assuming the baryon tilt
velocity remains constant before recombination, after which it
decays as the inverse scale factor. If dark matter decouples
at $z_{\mathrm{DM}}$, the ratio of baryon to dark-matter peculiar
velocities at last scattering will be
\begin{equation}
\left. \frac{|\vec{u}_{(\mathrm{b})}|}{|\vec{u}_{(\mathrm{c})}|}
\right|_{\mathrm{LSS}} \approx \frac{1+z_{\mathrm{DM}}}{1+z_{\mathrm{LSS}}} .
\label{eq:dmorb}
\end{equation}
We take dark-matter decoupling to be at redshift
\citep[see e.g.][]{loeb2005ssp}
\begin{equation}
z_{\mathrm{DM}} \sim \frac{10\, \mathrm{MeV}}{T_{\mathrm{CMB}} k_B} \left(\frac{M_{\sigma}}{\mathrm{100\, GeV}}\right) \left(\frac{M}{\mathrm{100\, GeV}}\right)^{1/4} \simeq 10^{11} ,
\end{equation} 
\noindent
where $k_B$ is Boltzmann's constant, $M_{\sigma}$ is the coupling
mass, $M$ is the particle mass, and the chosen values assume a
super-symmetric origin of the CDM particle. 

Unfortunately, the linearisation will break down at high redshift as
the expansion-normalized scales as $e^{-\alpha}$, so the value by
which the dark matter tilt is suppressed relative to the baryon tilt
is unclear. However, equation~(\ref{eq:dmorb}) strongly suggests that
the baryon--photon plasma dominates the momentum density at last
scattering and that its tilt should be properly determined at
$z_{\mathrm{LSS}}$ by equation~(\ref{eq:u-constraint}) with $\vec{P} =
(\rho_{\mathrm{b}} + 4 \rho_{\gamma}/3) \vec{u}_{(\mathrm{b})}$ on the
left-hand side. After this $\vec{u}_{(\mathrm{b})}$ decays as $1+z$ as
determined by (\ref{eq:vel-euler}). The usual procedure of assuming
that all components have the same tilt under-estimates
$|\vec{u}_{(\mathrm{b})}|$ at last scattering by the ratio of the
baryon--photon enthalpy to the total enthalpy. For the majority of our
results, we follow the usual procedure for consistency with previous
work.  We consider the effect of the improved velocity analysis in
Section \ref{sec:effect-impr-constr}, where we show that it will have
a significant impact on statistical studies, but does not change our
qualitative results.

\subsection{Parameters of the model}

For $\Lambda$CDM, the background model may be specified fully by the
physical densities in CDM ($\omega_{\mathrm{c},0}\equiv
\Omega_{\mathrm{c},0}h_{100}^2$ with $H_0 = 100 h_{100}
\,\mathrm{km}\,\mathrm{s}^{-1}\,\mathrm{Mpc}^{-1}$) and baryons
($\omega_{\mathrm{b},0}$) with $\Omega_{\Lambda,0}$ and
$\Omega_{K,0}$.  The Hubble constant and $\Omega_{\mathrm{M},0}$ are
then derived quantities.  Models with fixed $\omega_{\mathrm{c},0}$
and $\omega_{\mathrm{b},0}$ have the same early-universe history and
reproduce the same acoustic peak structure in the CMB spectra if the
angular-diameter distance to last-scattering and the primordial power
spectra are additionally held fixed~\citep{1999MNRAS.304...75E}.  The
Bianchi representation with structure constants in canonical form
further requires us to specify $e^{\alpha_0}$, although this is of no
physical consequence in the background.  In the perturbed model, the
current scale factor $e^{\alpha_0}$ determines the physical size over
which the shear eigenvectors rotate in space on a parallel-propagated
triad. For the simplified perturbed model, we must additionally
specify initial values for $\sigma_{12}$ and $\sigma_{13}$.  Due to
the rotational symmetry of the VII$_h$ structure constants about
$\vec{e}_1$, only $m=\pm 1$ anisotropies and polarization are
generated in this model, and are proportional to $[(\sigma_{12} \mp i
\sigma_{13})/H]_0$ respectively.
Varying the phase of $\sigma_{12} + i \sigma_{13}$ amounts to rotating
the sky about $\vec{e}_1$ (reflecting the residual freedom in the
choice of $\vec{e}_2$ and $\vec{e}_3$), while the
rotationally-invariant content depends on $\sigma_{12}^2 +
\sigma_{13}^2$.  We can, therefore, always choose $\sigma_{12} =
\sigma_{13}$ which we do for compatibility with previous studies.

The morphology of the CMB anisotropy and polarization patterns in the
Bianchi model is determined largely by the parameters
$\Omega_{\mathrm{M},0}$, $\Omega_{\Lambda,0}$ and the conformal Hubble
parameter $e^{\alpha_0} H_0$. The expansion-normalized shear
$(\sigma_{12}/H)_0$ and
$(\sigma_{13}/H)_0$ determine the amplitude.\footnote{%
The more refined treatment of tilt velocities requires one to specify also the
fraction of baryons to dark matter and the physical Hubble parameter. However,
the Doppler terms are generally only a small correction to the anisotropy
accumulated through the shear. The details of recombination introduce
further dependencies on the physical densities of baryons and dark matter.}
\citet{1973MNRAS.162..307C} denote the conformal Hubble parameter by
$x$, i.e.\
\begin{equation}
x=\dot{\alpha_0}e^{\alpha_0}=\sqrt{\frac{h}{\Omega_{K,0}}} ,
\end{equation}
where the latter relation arises directly from the FRW definition of
$\Omega_{K,0}$, with $K=-h$ in our case (Section \ref{sec:frw-limit}).
With $x$, $\Omega_{\mathrm{M},0}$ and $\Omega_{\Lambda,0}$ fixed, variations in
$e^{\alpha_0}$
change physical scales in the model (e.g.\ the age) but do not affect the
conformal properties. 
There is an approximate degeneracy amongst $\Omega_{\mathrm{M},0}$,
$\Omega_{\Lambda,0}$ and $x$ that preserves the morphology of
the Bianchi patterns~\citep{2006ApJ...644..701J,2007MNRAS.377.1473B}.
In our results, we follow~\citet{2006ApJ...644..701J}
by fixing $\omega_{\mathrm{c},0}$ and $\omega_{\mathrm{b},0}$ and use an ionization history
consistent with these choices. Further specifying $\Omega_{\mathrm{M},0}$ and
$\Omega_{\Lambda,0}$ determines $H_0$; the current scale factor is then
fixed by $x$.

\subsection{Summary of the calculation}

We assemble a hierarchy of multipole equations for $\Theta_{lm}$,
$E_{lm}$ and $B_{lm}$ using the results of
Section~\ref{sec:boltzmann-heirarchy}.
The Thomson scattering rate $\dot{\tau}$ requires a model for the
recombination (and potentially reionization) history, for which we use
\textsc{RECFAST}
\citep*{1999ApJ...523L...1S}.  

Starting at $z \simeq 1500$, the initial power is taken to be zero for
the polarization, with a pure dipole for the temperature arising from
the tilt of the baryon--photon fluid. Whilst the
universe remains optically thick, a small quadrupole term in the
temperature and polarization distributions arises from the equilibrium
between scattering and anisotropic redshifting due to shear; this is quickly
attained during the numerical integration and there is no requirement
to include it in the initial conditions. However, we verified that
starting significantly earlier ($z \simeq 1800$) made no difference to
the final results.

Since the power in the temperature modes declines very rapidly for
$l>15$, we truncate the hierarchy at $l=60$ without any special
boundary conditions. Performing the calculation with a higher
truncation ($l=120$) made no difference to the results for $l<30$.
During the numerical integration we ensure at each timestep $\delta
\eta \ll 1$, $\sigma \delta \eta \ll 1$ and $\dot{\tau}\delta \eta \ll
1$.

We investigate the polarization properties of the CMB in two models on
the degeneracy proposed by \cite{2006ApJ...644..701J}, namely
$(x,\Omega_{\Lambda,0},\Omega_{\mathrm{M},0})=(0.62,0,0.5)$ and $(1.0,
0.7, 0.2)$ respectively, both with ``right-handed'' parity. The latter
model is as close to a concordance value as the Bianchi fitting allows
(see Fig.\ 7 in \citealt{2007MNRAS.377.1473B}). In both cases, we take
$\omega_{\mathrm{b},0} = 0.022$ and $\omega_{\mathrm{c},0}=0.110$ and
the consistent recombination history with no reionization. These
models produce almost identical polarization patterns, for reasons
outlined below.  We briefly discuss the effects of altering the
ionization history in various ways (including reionization) in Section
\ref{sec:effect-recomb-model}.

In each case, we normalize such that the maximum temperature
anisotropy corresponds to $\Delta T = \pm 35\,\mu \mathrm{K}$. Note that the
amplitude of the polarization anisotropy simply scales linearly with
the magnitude of the temperature anisotropy.

\subsection{Results}

\begin{figure}
\includegraphics*[width=0.45\textwidth]{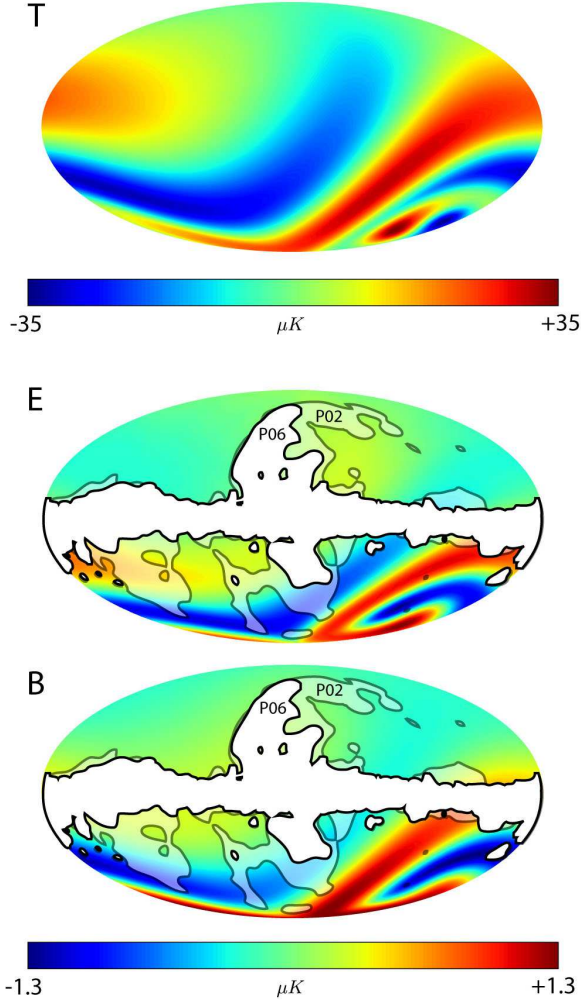}
\caption{Temperature (top), $E$-mode (middle) and $B$-mode
  polarization (bottom) maps for the Bianchi VII$_h$ model with
  $(x,\Omega_{\Lambda,0},\Omega_{\mathrm{M},0})$=$(0.62,0,0.5)$ and a
  consistent recombination history and no reionization. The maps have
  been transformed to the observational basis
  $(-\vec{p},\hat{\vec{e}}_{\theta},\hat{\vec{e}}_{\phi})$, which
  involves a parity change of the form~(\ref{eq:parity}), and rotated
  to match the orientation of the template given in
  \protect\cite{2006ApJ...644..701J}. The masks used in the
  \emph{WMAP} polarization analysis~\citep{2007ApJS..170..335P} are
  overlaid on the polarization maps.}\label{fig:pretty-maps}
\end{figure}

\begin{figure}
\begin{center}
  \includegraphics*[width=0.45\textwidth]{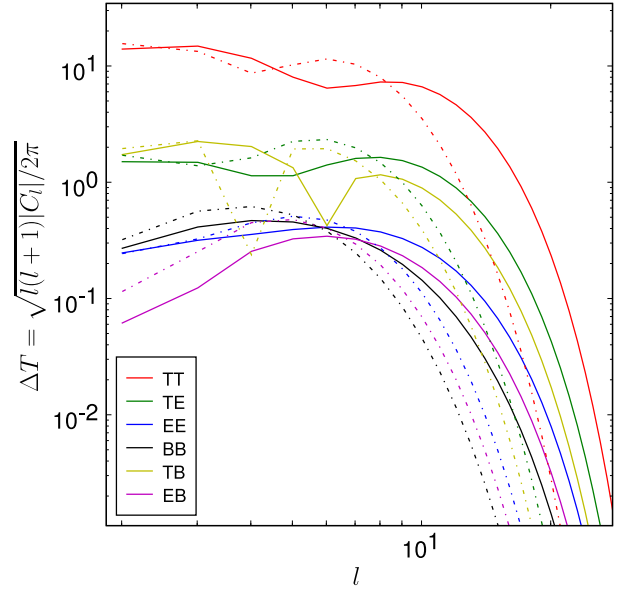}
\end{center}
\caption{Auto- and cross-correlation power spectra for
  the Bianchi models $(x,\Omega_{\Lambda,0},\Omega_{\mathrm{M},0})=(0.62,0,0.5)$ (solid
  lines) and $(x,\Omega_{\Lambda,0},\Omega_{\mathrm{M},0})=(1.0,0.7,0.2)$
  (dotted lines), normalized such that the maximum
$\Delta T = \pm 35\, \mu\mathrm{K}$. (The units of the vertical
axis are $\mu\mathrm{K}$.)
  The main difference between the models is a shift of power to larger scales
 (lower $l$) in the model with $\Lambda$;
  this is well understood in terms of the reduced focusing given lower
  $\Omega_{K,0}$ (see text), and causes no difference to our conclusions.
  Note that the $TB$ correlation is negative for $l<6$ and $l<5$ in the
  respective models.  }\label{fig:viih-model-comp}
\end{figure}

\begin{figure}
\begin{center}
\includegraphics*[width=0.45\textwidth]{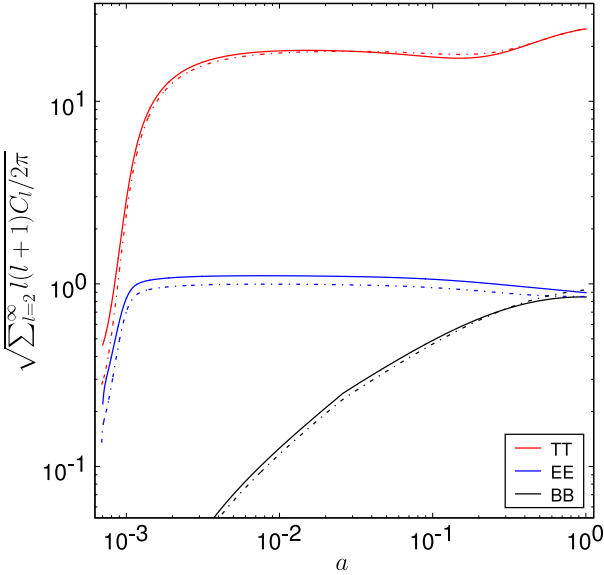}
\end{center}
\caption{Growth of observable r.m.s.\ signal in the $T$ and $E$- and $B$-mode
polarization plotted against
  $a=e^{\alpha-\alpha_0}=(1+z)^{-1}$ for the models
  $(x,\Omega_{\Lambda,0},\Omega_{\mathrm{M},0})$=$(0.62,0,0.5)$ (solid lines) and
  $(x,\Omega_{\Lambda,0},\Omega_{\mathrm{M},0})$=$(1.0,0.7,0.2)$ (dotted lines).
  Note that the power grows rapidly at high redshift while the shear
  is still significant, then remains constant (although it is
  transferred to higher $l$, which cannot be seen in this diagram). It
  is for this reason that the polarization is remarkably strong and relatively
insensitive to the cosmology along the line for which the 
VII$_h$ temperature patterns are degenerate.}\label{fig:viih-model-growth}
\end{figure}

\begin{figure}
  \begin{center}
    \includegraphics*[width=0.45\textwidth]{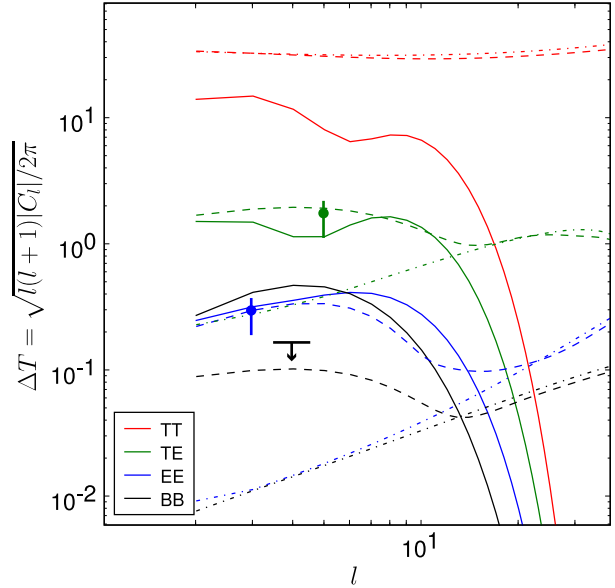}
  \end{center}
 \caption{Bianchi VII$_h$ induced power in the CMB (solid lines) for $(x,\Omega_{\Lambda,0},\Omega_{\mathrm{M},0})$=$(0.62,0,0.5)$ and no reionization, compared with Gaussian power from inhomogeneities for $(\omega_{\mathrm{c},0},\omega_{\mathrm{b},0},\sigma_8,r)=(0.110,0.022,0.7,0.3)$ with reionization optical depth $\tau=0$ (dash-dotted lines) and $\tau=0.10$ (dashed lines). The polarization data are from the \emph{WMAP}
   three-year release~\citep{2007ApJS..170..335P}.  From the $TE$ and
   $EE$ power spectra alone, the Bianchi-induced polarization can
   mimic the effect of early reionization in the standard scenario
   (the conventional interpretation of the large-scale polarization
   power seen by \emph{WMAP}).  However, the best-fit Bianchi model to
   the temperature map clearly over-produces $B$-mode power compared
   to the \emph{WMAP} upper limit (plotted) ruling out the simple
   model immediately.}
\label{fig:compare-concordance}
\end{figure}

The resulting temperature and polarization $E$- and $B$-mode maps for the
$\Omega_{\Lambda,0}=0$ case are illustrated in Fig.~\ref{fig:pretty-maps}. The level of the polarization is very high,
approximately $1\,\mu\mathrm{K}$.  Heuristically, this is because the shear
modes considered here decay as $(1+z)^3$, so that a substantial
portion of the final temperature anisotropy can be built up between
individual scattering events at high redshift.  Because of the
efficient conversion of $E$-modes to $B$,
(equation~\ref{eq:mix-terms}), the $B$-mode
contribution is of similar magnitude to the $E$-mode. 

Although computing the power spectra,
\begin{equation}
C_l^{XY} = \frac{1}{2l+1}\sum_m a^{X*}_{lm} a^{Y}_{lm},
\end{equation}
does throw away useful information in these
models, it provides a fast and efficient way to compare our results
with known, and robust, polarization constraints.
Since the multipole hierarchy does not
transfer power between different $m$ values, and the
implemented cosmology only generates anisotropies with $m=\pm 1$,
in forming the power spectrum we are throwing away only phase information.

Given the position of the Bianchi-like features on the sky given in
\cite{2006ApJ...644..701J}, we may be confident that the P06, and even
the P02, mask of the \emph{WMAP} polarization
analysis~\citep{2007ApJS..170..335P} could not hide the polarization
signal to a great extent (Fig.~\ref{fig:pretty-maps}).  Although the
relation between the Stokes parameters and $E$ and $B$ is non-local,
we note that maps of the Stokes parameters have their power localized
in a similar way to $E$ and $B$ (and $T$).  We therefore calculate the
full-sky power spectra without any consideration of the effect of
masking nor the weighting with the inverse of the (non-diagonal)
pixel-pixel noise covariance matrix that were employed by the
\emph{WMAP} team. Given that the r.m.s.\ Bianchi signal inside the
masks is lower than outside, we expect the effects of masking would
increase the estimated Bianchi power spectra over the full-sky values
plotted in Fig.~\ref{fig:viih-model-comp}.

The major difference between the two parameter sets considered
is that, for
the $\Omega_{\Lambda,0}=0.7$ case, the distinctive spiral pattern is
less `focused'. This is a well understood effect of reducing the
spatial curvature $\Omega_{K,0}$ to $0.1$ from its original value, $0.5$
(e.g.\ \citealt{1985MNRAS.213..917B}), and manifests itself as a shift of
power to
lower $l$ (see Fig.~\ref{fig:viih-model-comp}).
The existing statistical studies show that distinguishing these cases
observationally is currently not possible~\citep{2007MNRAS.377.1473B}.

There is no significant difference in the overall polarization power.
This follows because the majority of the power is built up rapidly at
high redshifts as the universe becomes optically thin and the shear
term has not decayed: at this point, the model is insensitive to the
values of $\Omega_{\Lambda,0}$ and $\Omega_{K,0}$ (Fig.~\ref{fig:viih-model-growth}).  Allowing $\omega_{\mathrm{b},0}$ to vary introduces
much more substantial variations in the relative level of
polarization; however, this introduces a further degree of freedom and
is beyond the scope of our current analysis. 

In Fig.~\ref{fig:compare-concordance}, we compare the power spectra
in the Bianchi $\Omega_{\Lambda,0}=0$ model with the power expected in a
`concordance' model with standard, statistically-isotropic and
homogeneous perturbations.
The latter spectra are computed using
CAMB \citep*{Lewis:1999bs}
%,2000ApJS..129..431Z}
for two models, one with no reionization (dot-dashed lines) and a
favoured reionization model (dashed lines; $\tau=0.10$).
Forming combined power spectra by adding the power from the
Bianchi model that best fits the temperature maps to that
from the concordance model
is inconsistent, since the models have different
parameters~\citep{2007MNRAS.377.1473B}.
However, ignoring
this problem and comparing the models as `templates' shows that, so
far as the $TE$ and $EE$ power spectra are concerned, the Bianchi model
can mimic the observed large-angle power that is conventionally attributed to
reionization. Of course, the `corrected' power in such a model would probably
lead to
an unfeasibly low estimate for $\tau$ in light of other data such
as the Gunn-Peterson constraints \cite*[e.g.][]{2006ARA&A..44..415F}. So,
at least with the fiducial simplified dynamics outlined in Section
\ref{sec:field-equations}, this already provides strong evidence
against the VII$_h$ model.

More challenging for the Bianchi model is the $B$-mode polarization, which is
at a similar level to the $E$-mode.  In
\citet{2007ApJS..170..335P}, the $B$-modes for $l<10$ are found to be
consistent with zero with errors better than
$\sigma \sim 0.1\,\mu\mathrm{K}^2$ at each multipole.
At this level, the signal-to-noise on the $B$-mode spectrum in
the Bianchi model should be at least unity for 
{\it each} $2<l<8$, and would have produced a highly significant detection
of large-angle $B$-modes overall.

Finally, the Bianchi models are not parity-invariant and one therefore
obtains a $TB$ and $EB$ cross-correlation
(Fig.~\ref{fig:viih-model-comp}). To get a rough estimate of the
current statistical power of these spectra in constraining the Bianchi
model, we compute the $\chi^2$ between the model prediction and the
\emph{WMAP} estimates of $C_l^{TE}$ and $C_l^{TB}$ available on the
LAMBDA website\footnote{\ttfamily http://lambda.gsfc.nasa.gov/}. We use
the spectra from $l=2$--16 and, since only the diagonal errors are
publically available, we ignore correlations between the estimates and
complications due to the shape of the low-$l$ likelihood. As noted
earlier, we also ignore the
effects of foreground masking and noise-weighting.\footnote{%
  We checked that the additional variance in the power spectrum
  estimates (when averaging over statistically-isotropic CMB
  fluctuations and noise) due to products between the two-point
  functions of the Bianchi signal and fluctuations is only a small
  correction to the errors computed by the \emph{WMAP} team.}  We find
reduced $\chi^2$ values of $0.5$ for $TB$ and $4.3$ for $EB$ for 15
degrees of freedom. The corresponding figures for null $C_l^{TB}$ and
$C_l^{EB}$ are $0.4$ and $0.6$. Note that, although the Bianchi power
is typically two orders of magnitude smaller in $EB$ than $TB$, the
$EB$ estimates have smaller errors as $C_l^T$ exceeds the variance of
the polarization noise on these scales. The interpretation of these
$\chi^2$ values is that the data are too noisy to distinguish the
Bianchi model from the null case for $TB$ (both are perfectly
consistent) but the $EB$ spectra disfavour the Bianchi model over the
null case.

\subsection{Effect of ionization history}\label{sec:effect-recomb-model}

Since the shear decays rapidly, $\sigma \propto (1+z)^3$, our
inclusion of a more detailed recombination calculation will affect the
temperature maps somewhat.  We take the $\Omega_{\Lambda,0}=0$ model
and run the Boltzmann hierarchy twice; first with instantaneous
recombination at $z=1100$ and then with the full \textsc{RECFAST}
history (with no reionization). There is no qualitative difference in
the temperature maps produced, but there is an approximately 15
percent decrease in the temperature amplitude in the latter case. Of
course, this is simply reflected in a slightly different estimate of
$(\sigma/H)_0$ and has no significant impact on previous probes of
Bianchi signatures.

However, the detailed recombination history does have a significant
impact on the amplitude of polarization. With the detailed model, the
amplitude is approximately five times larger than that derived from
the instantaneous model. Note that this puts the amplitude of
polarization in the instantaneous model in agreement with the
estimation in \citet{1968ApJ...153L...1R}. It is unsurprising that the
polarization is so sensitive to the recombination model, given that it
arises through the detailed interplay of the rapidly decreasing shear
and sharply peaked visibility function $\dot{\tau}e^{-\tau}$.

The effect of adding reionization is not as dramatic as for standard
FRW perturbations; this is because of the high level of the primordial
polarization relative to the temperature signal ($\sim 1/25$) on large
scales. Adding
reionization as early as $z=15$ produces only a $\sim$ 50 percent increase
in the $EE$ power.

\subsection{Effect of improved constraint model}\label{sec:effect-impr-constr}

\begin{figure}
  \includegraphics*[width=0.49\textwidth]{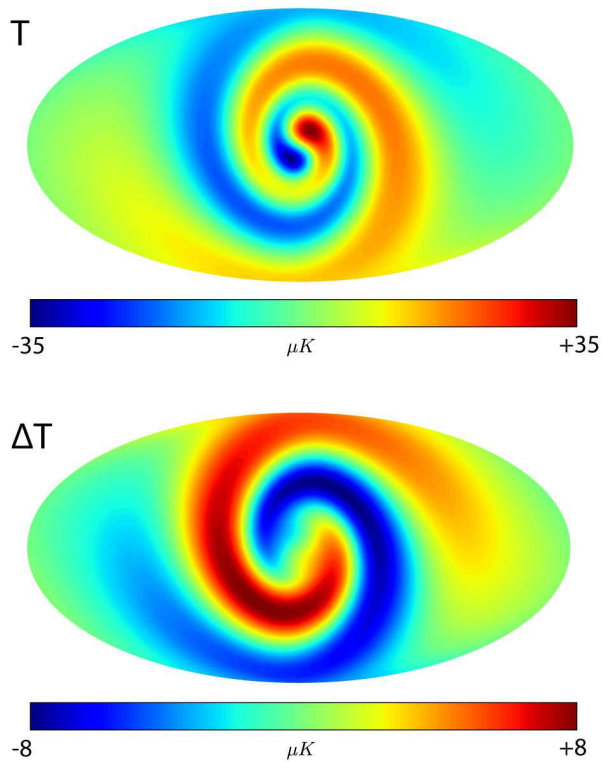}
\caption{(Top) Normalized temperature map for Bianchi VII$_h$ model
  $(x,\Omega_{\Lambda,0},\Omega_{\mathrm{M},0})$=$(0.62,0,0.5)$ with improved
  tilt constraint; (bottom) residuals in this map compared to the
  standard constraint map (Fig.~\ref{fig:pretty-maps}). The centre
  of the maps are here oriented down the $\vec{e}_1$ axis. }\label{fig:pretty-maps-bar}
\end{figure}

\begin{figure}
\begin{center}
  \includegraphics*[width=0.45\textwidth]{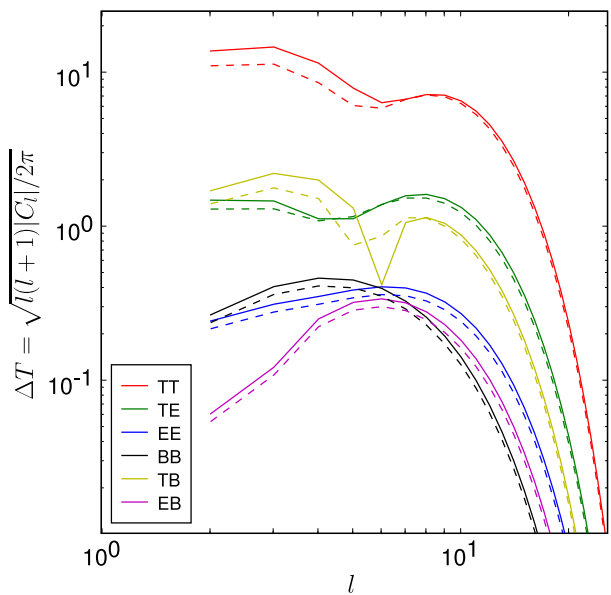}
\end{center}
\caption{Comparison of power spectra with standard constraint (solid lines)
  and improved constraint (dashed lines) after renormalizing the maximum
  temperature anisotropy to $\pm 35\,\mu \mathrm{K}$.
}\label{fig:viih-power-bar}
\end{figure}

The numerical results presented so far have derived the baryon tilt
assuming the same tilt for all particle species, consistent with
previous work. In this section, we analyze the effect of dropping this
assumption and adopting the more sophisticated model of
Section~\ref{sec:tilt-decay}.

One may see heuristically that the shear generally contributes more to
the temperature anisotropy than the dipole, because its integrated
effect (for models with matter domination over most of the line of
sight) scales as $\sim \Omega_{\mathrm{M},0}^{-1/2}(\sigma/H)_0
(1+z_{\mathrm{LSS}})^{3/2} \times \mathcal{O}(1)$ whereas the dipole
is imprinted instantaneously at scattering and scales in the improved
model as
\begin{equation}
|\vec{u}| \sim \frac{\sqrt{1+9h}(1+z_{\mathrm{LSS}})}{x \Omega_{\mathrm{b},0}} \left(\frac{\sigma}{H}\right)_0  .
\label{eq:modu}
\end{equation}
Using this new approach with the two sets of parameters considered
above, the ratio of the shear to Doppler contributions is $\sim 6$. If
instead we assume the same tilt for all species, as in the previous section,
we should replace $\Omega_{\mathrm{b},0}$ by $\Omega_{\mathrm{M},0}$
in equation~(\ref{eq:modu}) and this ratio becomes $\sim 30$. 

The effects of the improved treatment of tilt are thus twofold:
(a) an increase in the relative
level of the `distorted dipole' to the `distorted quadrupole'
component in the temperature maps; and (b) a decrease in the overall level
of polarization, given that the models are normalized to a fixed
maximum $\Delta T/T$, and only the quadrupole at high redshift is
responsible for producing polarization.

The revised temperature map and residual map for the
$(x,\Omega_{\Lambda,0},\Omega_{\mathrm{M},0})=(0.62,0,0.5)$ model are shown in
Fig.~\ref{fig:pretty-maps-bar}. The power spectra are plotted in
Fig.~\ref{fig:viih-power-bar}. As expected, the polarization
strength is somewhat lower after renormalization.
The effects are in accordance with our expectations:
the difference in the temperature maps amounts to a $15 \%$
effect, whilst the polarization level is reduced by approximately $10
\%$.  

It is clear that the details of how the tilt is treated will impact on
a detailed statistical comparison of the models with the \emph{WMAP} data, but
a full study is beyond the scope of the present
work.  However these effects are not sufficiently large to make the
Bianchi $B$-mode polarization unobservable at the three-year \emph{WMAP} sensitivity
(cf.\ Fig.~\ref{fig:compare-concordance}), or change our overall conclusions.

\section{Conclusions}\label{sec:conclusions}

We have derived the radiative-transfer equation for the CMB, including
polarization, in all nearly-FRW Bianchi universes in the form of a
hierarchy of multipole equations which can be easily integrated
numerically. These can be coupled with the dynamical (i.e.\ Einstein)
equations to compute maps of the CMB temperature anisotropies and
polarization in any such model. $B$-mode polarization is generic,
being produced in all Bianchi types except I.
We applied these equations to the Bianchi VII$_h$
case, with parameters tuned to address the anomalous features
observed in the CMB temperature on large
scales~\citep{2005ApJ...629L...1J,2006ApJ...644..701J}. 
Our treatment includes a more physical treatment of the tilt
velocity in CDM models with sub-dominant baryons.
Whilst this does not make a qualitative
difference to our results, more detailed statistical studies could
well be affected by its $\sim 20\%$ corrections.

Our temperature maps are similar to those derived from earlier
studies~\citep{1973MNRAS.162..307C,1985MNRAS.213..917B}, although the
amplitude is modified somewhat due to the better treatment of the
ionization history. Polarization maps, with the generality presented
here, do not appear to have been computed before.  Note also that for
these, a detailed treatment of recombination is required for accurate
results.  The power spectra of our type-VII$_{h}$ polarization maps
apparently put these models in contradiction of the large-scale
polarization results from \emph{WMAP} \citep{2007ApJS..170..335P}.

During the drafting of this paper, an analysis of universes equivalent
to Bianchi I models, tuned to account for the low CMB
quadrupole~\citep*{2006PhRvL..97t9903C}, was shown to give a similar level of
polarization to that computed here~\citep{2007astro.ph..2293C}.
This is not surprising given that the anisotropies
are tuned to address some of the same problems, and that the added
complications induced by the VII$_h$ geometry do not substantially
alter the amplitude of the effect (in the simplified dynamical model).
In the type-I model, the temperature anisotropy and $E$-mode polarization
are simply quadrupoles, and no $B$-modes are produced
(see equation~\ref{eq:mix-terms}). Although the type-I model
does not suffer the same observational constraints as type-VII$_h$
in polarization, the latter has the virtues in temperature of
resolving essentially all of the large-angle anomalies.

Our cursory glance over the available data appears to rule out the
VII$_h$ models employed in recent papers on the basis that they over-produce
$B$-mode power.
This is especially significant
given that the results hold for all models on the Bianchi degeneracy
line given in \cite{2006ApJ...644..701J}.  Our polarization
results, combined with the failure of the Bianchi degeneracy region to include
well-established values for the cosmological parameters, suggest that
the simple VII$_{h}$ model, as it stands, is
unsuitable to describe the available data.
However, to reject completely the hypothesis that our universe contains
anisotropic perturbations that are homogeneous under groups of motions
with Bianchi type VII$_{h}$ requires a fuller
treatment of the dynamics of the linearized model
(Section \ref{sec:field-equations}). We intend to address this problem,
and to search for statistical correlations between the morphology
of the generalized model's polarization and the \emph{WMAP} data, in future work.

\section*{Acknowledgments}
AP is supported by a STFC (formerly PPARC) studentship and scholarship
at St John's College, Cambridge. AC acknowledges a Royal Society
University Research Fellowship. We thank Kendrick Smith, Antony Lewis
and John Barrow for helpful discussions.

\bibliographystyle{mn2e} \bibliography{../refs.bib}
\end{document}